\documentclass[aps,twocolumn,pra,%nofootinbib,%
preprintnumbers,amsmath,amssymb,superscriptaddress]{revtex4}
\usepackage{graphicx}
\usepackage{color}

\definecolor{refkey}{rgb}{0.9451,0.2706,0.4941}
\definecolor{labelkey}{rgb}{0.9451,0.2706,0.4941}
\newcommand{\del} {\partial}
\newcommand{\eq} {equation}
\newcommand{\eqa} {eqnarray}
\newcommand{\NN} {\nonumber}
\begin{document}

\preprint{WIS/04/16-MAR-DPPA}

\title{Borel summability of perturbative series
in 4d $\mathcal{N}=2$ and 5d $\mathcal{N}=1$ theories}

\date{\today}

\author{Masazumi Honda}\email[]{masazumi.honda@weizmann.ac.il} 
\affiliation{{\it Department of Particle Physics, Weizmann Institute of Science, Rehovot 7610001, Israel}}

\begin{abstract}
We study weak coupling perturbative series 
in 4d $\mathcal{N}=2$ and 5d $\mathcal{N}=1$ supersymmetric gauge theories 
with Lagrangians.
We prove that
the perturbative series of these theories in zero instanton sector
are Borel summable for various observables.
Our result for 4d $\mathcal{N}=2$ case supports
an expectation from a recent proposal on a semiclassical realization of infrared renormalons in QCD-like theories,
where the semiclassical solution does not exist in $\mathcal{N}=2$ theories and
the perturbative series are unambiguous, namely Borel summable.
We also prove that
the perturbative series in arbitrary number of instanton sector
are Borel summable for a wide class of theories.
It turns out that
exact results can be obtained by 
summing over the Borel resummations in each number of instanton sector.
\end{abstract}

\maketitle

%\tableofcontents

%%%%%%%%%%%%%%%%%%%%%%%%%%%%%%%%%%%%%%%%%%%%%%%%%%%%%%

\noindent

%%%%%%%%%%%%%%%%%%%%%%%%%%%%%%%%%%%%%%%%%
%%%%%%%%%%%%%%%%%%%%%%%%%%%%%%%%%%%%%%%%%
%%%%%%%%%%%%%%%%%%%%%%%%%%%%%%%%%%%%%%%%%
\section{Introduction}
%%%%%%%%%%%%%%%%%%%%%%%%%%%%%%%%%%%%%%%%%
%%%%%%%%%%%%%%%%%%%%%%%%%%%%%%%%%%%%%%%%%
%%%%%%%%%%%%%%%%%%%%%%%%%%%%%%%%%%%%%%%%%
Weak coupling perturbation theory in quantum field theory (QFT)
typically yields asymptotic series \cite{Dyson:1952tj}.
One of standard ways to resum asymptotic series
is Borel resummation:
when we have a perturbative series $I(g)=\sum_{\ell =0}^\infty c_\ell g^{a+\ell}$,
Borel summation is given by
\begin{\eq}
\int_0^\infty dt\ e^{-\frac{t}{g}}  \mathcal{B}I(t) ,
\label{eq:Borel}
\end{\eq}
where $\mathcal{B}I(t)$ is 
analytic continuation of 
the Borel transformation  
$\sum_{\ell =0}^\infty \frac{c_\ell}{\Gamma (a+\ell )} t^{a+\ell -1}$
after taking the summation.
However, 
it is expected that
perturbative series in typical interacting QFT is not
Borel summable 
due to singularities at positive real axis 
in $\mathcal{B}I(t)$
\cite{'tHooft:1977am}.

Nevertheless
one might wonder
if situation becomes better in theories with huge symmetries
such as extended supersymmetric (SUSY) theories.
Indeed some SUSY theories
have some protected quantities,
which do not receive quantum corrections or
receive only one-loop correction.
On the other hand
it is totally unclear for unprotected observables,
which receive higher loop perturbative corrections 
and non-perturbative corrections,
when perturbative series are convergent, Borel summable or
non-Borel summable.

Recently
there appeared an interesting hint \cite{Argyres:2012vv}
to this question
from resurgence approach 
to asymptotic free gauge theory on $S^1 \times \mathbb{R}^3$
(see also \cite{Dunne:2012ae}).
In \cite{Argyres:2012vv} the authors 
have proposed a semiclassical realization of infrared (IR) renormalons,
which correspond to leading singularities in Borel plane
of QCD-like theories.
In this proposal,
the IR renormalons are realized by 
so-called bion-anti-bion contributions \cite{Unsal:2007jx}
and cancel ambiguities of resummation in perturbatation theory.
This mechanism is QFT analogue of Bogomolny-Zinn-Justin mechanism 
in quantum mechanics \cite{Bogomolny:1980ur}.
While this scenario is expected 
for non-SUSY theory and 
$\mathcal{N}=1$ 
theory, 
there are 
no bion solutions 
in theories with $\mathcal{N}\geq 2$ SUSY \cite{Poppitz:2011wy}.
Therefore
it is expected that
perturbative expansions themselves in $\mathcal{N}\geq 2$ theories
do not have ambiguities and 
are Borel summable.
However
there are very few examples to support this expectation so far,
which are $S^4$ partition functions and SUSY Wilson loops 
in $SU(2),U(2)$ gauge theories
\cite{Russo:2012kj,Aniceto:2014hoa},
and extremal correlators in superconfomal QCD 
with $SU(2)$ and $SU(3)$ gauge groups \cite{Gerchkovitz:2016gxx} (see also \cite{Baggio:2014sna}).

Main purpose of this paper is
to provide strong evidence for this expectation.
Namely we prove Borel summability \footnote{
By ``Borel summability",
we mean Borel summability along positive real axis throughout this paper.
However, 
if the localization formula is still correct for complex $g$,
then our argument also shows Borel summability in right-half complex plane
except squashed sphere partition functions.
} 
of perturbative series 
in general 4d $\mathcal{N}=2$ theories with Lagrangians
for various observables \footnote{
Strictly speaking,
we prove that
the perturbative series are convergent or
asymptotic but Borel summable.
Hence our result is consistent 
with expected convergence in planar limit \cite{'tHooft:1982tz}.
We also assume that
$S^4$ partition function is well defined.
Physically this corresponds to
conformal field theory, its mass deformation and asymptotic free theory.
}.
Our main tool is
the localization method \cite{Pestun:2007rz},
which reduces
path integrals to finite dimensional integrals for a class of observables.
We also prove 
Borel summability
in arbitrary number of instanton sector
when we know explicit expressions of 
instanton conrrections to localization formula,
which are captured by 
so-called Nekrasov instanton partition function \cite{Nekrasov:2002qd}.
In section \ref{sec:4d}
we give proofs
for $S^4$ partition function, SUSY Wilson loop,
Bremsstrahrung function, extremal correlator and
squashed $S^4$ partition function. 
As a conclusion
our result strongly supports
the expectation from the proposal \cite{Argyres:2012vv} 
on the semiclassical realization of IR renormalons in QCD-like theories. 

Another main result of this paper is that
the Borel resummation in a fixed number of instanton sector
is exactly the same as the truncation of the full result to this sector.
This means that
perturbative series in each number of instanton sector
does not mix with each other 
from the viewpoint of resummation.
This feature is quite different 
from successful examples of resurgence approach,
where perturbative expansion in zero instanton sector 
is related to ones in non-zero instanton sector.
We discuss this point in more detail in sec.~\ref{sec:discussions}.

In section \ref{sec:5d}
we also discuss 5d $\mathcal{N}=1$ gauge theories.
While 5d gauge theory is not renormalizable in the sense of power counting,
one can reasonably compute 
some observables in UV complete theories \cite{Seiberg:1996bd}.
We can prove 
Borel summatbility of perturbative series
for squashed $S^5$ partition function and SUSY Wilson loop.

%%%%%%%%%%%%%%%%%%%%%%%%%%%%%%%%%%%%%%%%%
%%%%%%%%%%%%%%%%%%%%%%%%%%%%%%%%%%%%%%%%%
%%%%%%%%%%%%%%%%%%%%%%%%%%%%%%%%%%%%%%%%%
\section{4d $\mathcal{N}=2$ theory}
\label{sec:4d}
%%%%%%%%%%%%%%%%%%%%%%%%%%%%%%%%%%%%%%%%%
%%%%%%%%%%%%%%%%%%%%%%%%%%%%%%%%%%%%%%%%%
%%%%%%%%%%%%%%%%%%%%%%%%%%%%%%%%%%%%%%%%%
%%%%%%%%%%%%%%%%%%%%%%%%%%%%%%%%%%%%%%%%%
%%%%%%%%%%%%%%%%%%%%%%%%%%%%%%%%%%%%%%%%%
\subsection{$S^4$ partition function}
%%%%%%%%%%%%%%%%%%%%%%%%%%%%%%%%%%%%%%%%%
%%%%%%%%%%%%%%%%%%%%%%%%%%%%%%%%%%%%%%%%%
We begin with partition function on $S^4$.
Let us consider
4d $\mathcal{N}=2$ theory 
with semi-simple gauge group $G=G_1 \times\cdots\times G_n$ and
$N_f$ hyper multiplets 
of representations $(\mathbf{R_1} ,\cdots ,\mathbf{R_{N_f}} )$.
Thanks to the localization method,
the $S^4$ partition function can be
expressed in terms of 
the following finite dimensional integral \cite{Pestun:2007rz}
\begin{\eq}
Z_{S^4}
= \int_{-\infty}^\infty d^{|G|} a\ 
Z_{\rm VdM} Z_{\rm cl} Z_{\rm 1loop} Z_{\rm inst}  ,
\label{eq:ZS4}
\end{\eq} 
where $a$ takes values in Cartan subalgebra of $G$.
$Z_{\rm cl}$ and $Z_{\rm 1loop}$ are
classical and one-loop contributions around localization locus, respectively
\footnote{
We can also include FI-term and mass.
Effects of the FI-term and mass are addition of 
a linear function of 
$a$ in the exponent of $Z_{\rm cl}$ and
a constant shift in one-loop and instanton parts, respectively.
These do not spoil our proof
as long as the mass is real.
}:
\begin{\eqa}
&& 
Z_{\rm VdM}(a)
= \prod_{\alpha \in {\rm root}_+ } (\alpha\cdot a)^2  ,\NN\\
&& Z_{\rm cl}(a)
= \exp{\Bigl[ -\sum_{p=1}^n \frac{1}{g_p} {\rm tr} (a^{(p)} )^2  \Bigr]} ,\NN\\
&& Z_{\rm 1loop}(a) 
= \frac{\prod_{\alpha \in {\rm root}_+ }H^2 (\alpha\cdot a ) }
   {\prod_{m=1}^{N_f} \prod_{\rho_m \in \mathbf{R_m}} H( \rho_m \cdot a )} ,\NN\\
&& H(x) = e^{-(1+\gamma )x^2} G(1+ix) G(1-ix) ,
\end{\eqa} 
where the parameter $g_p$ is proportional
to square of Yang-Mills coupling of the gauge group $G_p$,
$\rho_m$ is weight vector of the representation $\mathbf{R_m}$,
$\gamma$ is Euler constant
and $G(x)$ is Barnes $G$-function.
$Z_{\rm inst}$ is contributions from instantons,
which are described 
by Nekrasov instanton partition function $Z_{\rm Nek}(a)$ \cite{Nekrasov:2002qd}
with $\Omega$-background parameters $\epsilon_1 =\epsilon_2 =1$ \footnote{
We are considering unit $S^4$.
For $S^4$ with radius $r$, 
the $\Omega$-background parameters are $\epsilon_1 =\epsilon_2 =r^{-1}$.
}
as
\begin{\eq}
Z_{\rm inst}(a) = |Z_{\rm Nek}(a)|^2 
= \sum_{ \{ k_p \}  =0}^\infty e^{-\sum_{p=1}^n \frac{k_p}{g_p}} Z_{\rm inst}^{(k_1 ,\cdots ,k_n )} (a) .
\end{\eq}
Now we are interested in 
weak coupling expansion of $Z_{S^4}$ 
in a fixed number of instanton sector:
\begin{\eq}
Z_{S^4}^{(k_1 ,\cdots ,k_n)}(g)
=  \int_{-\infty}^\infty d^{|G|} a\
Z_{\rm VdM} Z_{\rm cl} Z_{\rm 1-loop} Z_{\rm inst}^{(k_1 ,\cdots ,k_n)}  .
\end{\eq} 
This has the following weak coupling expansion:
\begin{\eq}
Z_{S^4}^{(k_1 ,\cdots ,k_n)} 
\sim \sum_{ \{ \ell_p \} =0}^\infty  
c_{\ell_1 , \cdots ,\ell_n}^{(k_1 ,\cdots ,k_n)}
 \prod_{p=1}^n g_p^{\frac{{\rm dim}(G_p )}{2} +\ell_p} .
\end{\eq}
Here we prove that
the small-$g$ expansion of $Z_{S^4}^{(k_1 ,\cdots ,k_n )}(g)$ is Borel summable 
and this is true also for various observables.
It is also interesting to ask 
property of the expansion by instanton number
as in \cite{Aniceto:2014hoa}
but this is beyond scope of this paper.

%%%%%%%%%%%%%%%%%%%%%%%%%%%%%%%%%%%%%%%%%
\subsubsection*{$SU(N)$ superconformal QCD}
%%%%%%%%%%%%%%%%%%%%%%%%%%%%%%%%%%%%%%%%%
First 
we focus on the 4d $\mathcal{N}=2$ $SU(N)$ superconformal QCD 
with 
$2N$ fundamental hyper multiplets.
We will consider more general theory later.
The $S^4$ partition function of this theory 
in $k$-instanton sector is given by
\begin{\eqa}
Z_{\rm SQCD}^{(k)}
&=&  \int_{-\infty}^\infty d^N a\ \delta \Bigl( \sum_j a_j \Bigr) \prod_{i<j}(a_i -a_j )^2 \NN\\
&& e^{-\frac{1}{g} \sum_j a_j^2} 
\frac{\prod_{i<j}H^2 (a_i -a_j ) }{\prod_j H^{2N}(a_j )}  Z_{\rm inst}^{(k)} (a) ,
\label{eq:SQCD}
\end{\eqa} 
where the delta function comes from speciality of $SU(N)$.
We would like to know property of small-$g$ expansion of $Z_{\rm SQCD}^{(k)}$.
For this purpose let us take the coordinate
\begin{\eq}
a_i = \sqrt{t} \hat{x}_i ,
\end{\eq}
where $\hat{\mathbf{x}} =(\hat{x}_1 ,\cdots ,\hat{x}_N )$ is the unit vector
spanning unit $S^{N-1}$.
Then we rewrite the partition function as \footnote{This expression for $SU(2)$ already appeared in \cite{Russo:2012kj}.
}
\begin{\eqa}
 Z_{\rm SQCD}^{(k)}
=  \int_0^\infty dt\ e^{-\frac{t}{g}} f^{(k)} (t) ,
\label{eq:LaplaceSQCD}
\end{\eqa} 
where
\begin{\eqa}
&& f^{(k)} (t) 
= \frac{t^{\frac{N^2 -3}{2}}}{2}
     \int_{S^{N-1}} d^{N-1}\hat{x}\ \delta \Bigl( \sum_j \hat{x}_j \Bigr)  
 h^{(k)} (t,\hat{x}) ,\NN\\
&& h^{(k)} (t,\hat{x}) 
= Z_{\rm VdM}(\hat{x})
Z_{\rm 1-loop}( \sqrt{t} \hat{\mathbf{x}})
 Z_{\rm inst}^{(k)} (\sqrt{t}\hat{\mathbf{x}})  .
\end{\eqa}
Note that
this takes the form of the Laplace transformation
as in the Borel resummation formula \eqref{eq:Borel}.
So it would be natural to expect that
$f^{(k)}(t)$ is Borel transformation of the original perturbative series,
namely analytic continuation of $\sum_{\ell =0}^\infty \frac{c_\ell^{(k)} }{\Gamma (\frac{N^2 -1 +\ell}{2})} t^\ell$ \footnote{
Below we simply refer to analytic continuation of formal Borel transformation
as Borel transformation.
}.
We will prove this in the following steps:
(I) 
We show that
the integrand $h^{(k)}(t,\hat{x})$ is 
identical to analytic continuation of a convergent power series of $t$.
(II) 
We ask if we can exchange 
the power series expansion of $h^{(k)}(t,\hat{x})$ and integration over $\hat{x}$.
We show this by proving uniform convergence of the small-$t$ expansion.
(III) 
The Laplace transformation \eqref{eq:LaplaceSQCD}
guarantees that 
the coefficient of the perturbative series of $f^{(k)}(t)$ 
at $\mathcal{O}(t^{\frac{N^2 -3}{2}+\ell} )$ 
is given by $c_\ell^{(k)} /\Gamma (\frac{N^2 -1 +\ell}{2})$.

For simplicity
we first focus on zero-instanton sector
and
non-zero instanton sector will be considered later.
By using product representation of the Barnes $G$-function \footnote{
$G(1+z)
=(2\pi )^{\frac{z}{2}} e^{-\frac{(1+z+\gamma z^2)}{2}}
\prod_{n =1}^\infty \left( 1+\frac{z}{n} \right)^n e^{-z +\frac{z^2}{2n}}$.
},
$h^{(0)}(t,\hat{x})$ can be written as
\begin{\eq}
h^{(0)}(t,\hat{x}) 
=
Z_{\rm VdM}(\hat{x})
\prod_{n=1}^\infty \frac{\prod_{i<j} \left( 1+\frac{t (\hat{x}_i -\hat{x}_j)^2 }{n^2} \right)^{2n}}
{\prod_j \left( 1+\frac{t (\hat{x}_j )^2}{n^2} \right)^{2Nn}} .
\label{eq:h_product}
\end{\eq}
Plugging 
$\sum_{n=1}^\infty n\log{\left( 1+\frac{x^2}{n^2} \right)}
=- \sum_{\ell =1}^\infty 
 \frac{(-1)^\ell \zeta (2\ell -1) }{\ell}  x^{2\ell}$
into this,
we find
generating function for the small-$t$ expansion of $h^{(0)}$ as
\begin{\eqa}
&& Z_{VdM}(\hat{x})
\exp \Biggl[ 
-2\sum_{i<j }\sum_{\ell =2}^\infty  \frac{(-t)^\ell \zeta (2\ell -1) }{\ell} (\hat{x}_i -\hat{x}_j ) ^{2\ell} \NN\\
&&\ \ +2N \sum_{j }\sum_{\ell =2}^\infty \frac{(-t)^\ell \zeta (2\ell -1) }{\ell}  \hat{x}_j^{2\ell}
\Biggr] .
\label{eq:gen}
\end{\eqa}
The small-$t$ expansion of $h^{(0)}$ has a finite radius of convergence $t_0$,
which is dependent on $\hat{x}$ but larger than $1$.
Hence $h^{(0)}$ is 
the same as analytic continuation of the convergent power series of $t$.

Next 
we show commutativity of the summation and integration over $\hat{x}$
by proving uniform convergence of the small-$t$ expansion of $h^{(0)}$.
For this purpose
it is convenient to apply Weierstrass's M-test.
Namely we find a sequence $\{ M_\ell \}$ satisfying
$| h^{(0)}_\ell (\hat{x})| < M_\ell $ and 
$\sum_{\ell =0}^\infty M_\ell  <\infty $
for fixed $t$.
We can easily find such a series.
For example,
since $\zeta (2\ell -1)<2,\ |\hat{x}|\leq 1$,
a generating function $\bar{h}^{(0)}(t)$ of $M_\ell$ can be obtained 
by the replacements 
$(-1)^{\ell +1}\zeta (2\ell -1)(\hat{x}_i -\hat{x}_j )^{2\ell}\rightarrow 2$,
$(-1)^{\ell }\zeta (2\ell -1) \hat{x}_j^{2\ell}\rightarrow 2$
in \eqref{eq:gen}:
\begin{\eq}
 \bar{h}^{(0)}(t)  
= \exp \Biggl[ 
2N(3N-1)\sum_{\ell =2}^\infty  \frac{t^\ell  }{\ell} 
\Biggr]
= \left( \frac{e^{-t}}{1-t} \right)^{2N(3N-1)} .
\label{eq:Mtest}
\end{\eq}
Thus $f^{(0)}(t)$ is actually
the Borel transformation of 
the original perturbative series.

To show Borel summability of the perturbative series
we should ask analytic property of the Borel transformation $f^{(0)}(t)$.
First $f^{(0)}(t)$ does not have branch cut along $t\in\mathbb{R}_+$
though this has a branch cut 
along $t\in\mathbb{R}_-$
for even $N$ \footnote{
We are taking branch cut of $\sqrt{z}$ in negative real axis.
}.
Structure of poles can be easily seen 
from the infinite product representation \eqref{eq:h_product} of $h^{(0)}$.
Since $h^{(0)}$ cannot have poles 
unless $t\in\mathbb{R}_-$
and integration is finite dimensional over compact region,
$f^{(0)}$ does not have poles along positive real axis.
Thus, we conclude that
the perturbative series of $Z_{\rm SQCD}$ in zero instanton sector is Borel summable.
As we discuss below,
generalization to other theories and observables, and
nonzero instanton sector are quite straightforward.

%%%%%%%%%%%%%%%%%%%%%%%%%%%%%%%%%%%%%%%%%
\subsubsection*{General theory with Lagrangian}
%%%%%%%%%%%%%%%%%%%%%%%%%%%%%%%%%%%%%%%%%
Generalization to other theories is quite parallel
to the case of the SQCD.
First we insert delta function constraint $\Delta (a)$ 
to the integrand as in \eqref{eq:SQCD}
such that the following coordinate
\begin{\eq}
a_i^{(p)} = \sqrt{t_p} \hat{x}_i^{(p)} ,
\label{eq:trans}
\end{\eq}
parametrizes sphere with the radius $\sqrt{t_p}$.
Then the partition function again takes the form of Laplace transformation
with multi variables:
\begin{\eqa}
 Z^{(k_1 ,\cdots ,k_n )}
=  \int_0^\infty d^n t\ 
 e^{-\sum_{p=1}^n \frac{t_p}{g_p}} f^{(k_1 ,\cdots ,k_n )}(t) ,
\label{eq:Laplace}
\end{\eqa} 
where
\begin{\eqa}
&& f^{(k_1 ,\cdots ,k_n )} (t) 
=\frac{t^{\frac{{\rm dim}(G)}{2}-1}}{2^n}  
\int_{\rm sphere}d\hat{x}\ 
\Delta (\hat{x}) h^{(k_1 ,\cdots ,k_n )} (t,\hat{x}) ,\NN\\
&& h^{(k_1 ,\cdots ,k_n )} (t) 
=\left. Z_{\rm VdM}(\hat{x}) Z_{\rm 1loop} Z_{\rm inst}  
\right|_{a_i^{(p)} = \sqrt{t_p} \hat{x}_i^{(p)}} ,
\end{\eqa}
with $t^{\frac{{\rm dim}(G)}{2}-1} 
=\prod_{p=1}^n t_p^{\frac{{\rm dim}(G_p )}{2}-1}$.
Let us focus on zero-instanton sector again.
Then we can always prove that
$h^{(k_1 ,\cdots ,k_n )}$ with $k_p =0$ gives
a uniform convergent power series of $t$.
Namely we can always construct 
convergent series as in \eqref{eq:Mtest}
to pass Weierstrass's M-test.
Hence $f^{(k_1 ,\cdots ,k_n )} (t)$ at zero instanton sector
is actually Borel transformation.
The Borel transformation cannot have poles and branch cut 
along positive real axis.
Therefore the perturbative series of $Z_{S^4}$ in zero instanton sector 
is Borel summable for general $\mathcal{N}=2$ theory with Lagrangian.

%%%%%%%%%%%%%%%%%%%%%%%%%%%%%%%%%%%%%%%%%
\subsubsection*{Nonzero instanton sector}
%%%%%%%%%%%%%%%%%%%%%%%%%%%%%%%%%%%%%%%%%
Generalization to arbitrary number of instanton sector
is also straightforward
when we know explicit forms of Nekrasov partition functions.
This is
because $Z_{\rm inst}^{(k_1 ,\cdots ,k_n )}(a)$ for all the known cases 
is rational function of $a$,
whose poles are not located at real axis.
For example,
Nekrasov partition function 
of $U(N)$ SQCD with $N_f$ fundamentals and anti-fundamentals is 
given by \footnote{To get $SU(N)$,
we shall strip decoupling $U(1)$ part \cite{Alday:2009aq}.}
\begin{\eq}
Z_{\rm Nek,SQCD}(a)
=\sum_{\{ Y_1 ,\cdots ,Y_N \}} q^{|Y|}
\frac{\left( \prod_{j=1}^N  n^f_j (a,Y)  \right)^{N_f} }
     {\prod_{i,j =1}^N n^V_{i,j} (a ,Y ) } ,
\end{\eq}
where $Y_j$ is Young diagram associated with $a_j$ and 
\begin{\eqa}
&& n^V_{i,j} (a,Y )
= \prod_{s\in Y_i} E_{ij}(a ,s)( \epsilon_1 +\epsilon_2 -E_{ij}(a ,s) ) ,\NN\\
&& E_{ij}(a ,s)
= -\epsilon_1 A_{Y_j} (s) +\epsilon_2 ( L_{Y_i} (s) +1) -i(a_i -a_j ) ,\NN\\
&& n^f_j (a,Y) 
= \prod_{s\in Y_j} \phi_j (a,s) (\phi_j (a,s) +\epsilon_1 +\epsilon_2 )   ,\NN\\
&& \phi_j (a,s) = -ia_j  +\epsilon_1 (s_h -1) +\epsilon_2 (s_v -1) .
\end{\eqa}
Here $s=(s_h ,s_v )$ labels
box in Young tableau at $s_h$-th column and $s_v$-th row,
and $L_Y$ ($A_Y$) is leg (arm) length of Young tableau $Y$ at $s$.
Although the contribution from each Young diagram
may have poles along real axis of $a$,
poles of this type are canceled 
after summing over Young diagrams with the same instanton number \cite{Pestun:2007rz}.
This is true 
unless $m_1 \epsilon_1 +m_2 \epsilon_2$ ($m_{1,2} \in\mathbb{Z}$)
can be purely imaginary
but $Z_{S^4}$
for this case is ill-defined and
therefore we do not consider this case.
This feature is common among all the cases,
where 
expressions of Nekrasov partition function are explicitly known.

Keeping this in mind,
now we can easily prove Borel summability
as in zero instanton sector.
Since $h^{(k_1 ,\cdots ,k_n )}$ is
product of $h^{(0 ,\cdots ,0)}$ and rational function of $\sqrt{t_p}$,
small-$t$ expansion of $h^{(k_1 ,\cdots ,k_n )}$ is always uniform convergent 
and
hence $f^{(k_1 ,\cdots ,k_n )} (t)$ is always Borel transformation
of the original perturbative series.
We can easily see that
the Borel transformation
cannot have poles and branch cuts along positive real axis
by using the above property of $Z_{\rm inst}$. 
Thus 
the perturbative series of $Z_{S^4}$ in arbitrary number of instanton sector 
is Borel summable.

%%%%%%%%%%%%%%%%%%%%%%%%%%%%%%%%%%%%%%%%%
%%%%%%%%%%%%%%%%%%%%%%%%%%%%%%%%%%%%%%%%%
\subsection{Other observables}
%%%%%%%%%%%%%%%%%%%%%%%%%%%%%%%%%%%%%%%%%
%%%%%%%%%%%%%%%%%%%%%%%%%%%%%%%%%%%%%%%%%
%%%%%%%%%%%%%%%%%%%%%%%%%%%%%%%%%%%%%%%%%
\subsubsection*{Supersymmetric Wilson loop}
%%%%%%%%%%%%%%%%%%%%%%%%%%%%%%%%%%%%%%%%%
Generalization to some other observables is straightforward as well.
First let us consider the Wilson loop
\begin{\eq}
W_{\mathbf{R}}(C)
={\rm tr}_{\mathbf{R}}
P\exp{\Biggl[ i\oint_C ds (A_\mu \dot{x}^\mu +i\Phi ) \Biggr]} ,
\label{eq:Wilson}
\end{\eq}
where $\Phi$ is the adjoint scalar in vector multiplet.
The Wilson loop is supersymmetric 
when the contour $C$ is the grand circle of $S^4$. 
By applying the localization method,
VEV of the Wilson loop is represented 
by the following VEV of the matrix model
\begin{\eq}
\langle W_{\mathbf{R}}({\rm Circle})  \rangle
=\langle {\rm tr}_{\mathbf{R}} e^a \rangle_{\rm M.M.} .
\end{\eq}
Since this is just finite linear combination of
exponentials,
this does not give anything harmful.
Thus 
repeating the above arguments,
we can prove that
perturbative series of the SUSY Wilson loop is Borel summble.
Obviously 
products of Wilson loops also give Borel summable perturbative series.

%%%%%%%%%%%%%%%%%%%%%%%%%%%%%%%%%%%%%%%%%
\subsubsection*{Bremsstrahrung function in SCFT}
%%%%%%%%%%%%%%%%%%%%%%%%%%%%%%%%%%%%%%%%%
Bremsstrahrung function $B$
appears in cusp anomalous dimension of small boost:
$\Gamma_{\rm cusp}(\varphi ) =B\varphi^2 +\mathcal{O}(\varphi^4 )$,
where $\varphi$ is the boost parameter.
This determines an energy radiated by accelerating quarks \cite{Correa:2012at}.
It was conjectured that
the Bremsstrahrung function in $\mathcal{N}=2$ superconformal theory
is given 
by the following VEV of the matrix model \cite{Fiol:2015spa}
\begin{\eq}
B=\frac{1}{4\pi^2}\left. \frac{\del}{\del b}
\langle {\rm tr} e^{ba} \rangle_{\rm M.M.} \right|_{b=1} ,
\end{\eq}
which is formally derivative of the supersymmetric Wilson loop in fundamental representation
with winding number $b$.
Since we have shown Borel summability for the Wilson loop,
$B$
is also Borel summable.

%%%%%%%%%%%%%%%%%%%%%%%%%%%%%%%%%%%%%%%%%
\subsubsection*{Extremal correlator in SCFT}
%%%%%%%%%%%%%%%%%%%%%%%%%%%%%%%%%%%%%%%%%
Next we consider the correlation function
\begin{\eq}
\left\langle
\mathcal{O}_{I_1}(x_1 ) \cdots \mathcal{O}_{I_1}(x_n )
\overline{\mathcal{O}}_{\bar{J}} (y)
\right\rangle ,
\end{\eq}
where $\mathcal{O}_I$ and $\overline{\mathcal{O}}_{\bar{I}}$ are 
chiral and anti-chiral primary operators, respectively.
This is often called extremal correlator.
It is known that
this is determined by the two point function 
$\left\langle
\mathcal{O}_{I}(x ) \overline{\mathcal{O}}_{\bar{J}} (y) 
\right\rangle $.
It was shown \cite{Gerchkovitz:2016gxx} that
the two point function is 
given by a ratio of (finite) linear combination of 
the quantity
$\langle \prod_j ({\rm tr}a^{m_j})^{n_j} \rangle_{\rm M.M.}$.
We can show Borel summability for this quantity by repeating the above aruments 
and 
hence perturbative series for the extremal correlator is also Borel summable.

%%%%%%%%%%%%%%%%%%%%%%%%%%%%%%%%%%%%%%%%%
\subsubsection*{Squashed $S^4$ partition function}
%%%%%%%%%%%%%%%%%%%%%%%%%%%%%%%%%%%%%%%%%
Next let us consider partition function $Z_{S_b^4}$ on squashed $S^4$
with a squashing parameter $b$.
This has a simple relation to 
supersymmetric Renyi entropy \cite{Nishioka:2013haa,Crossley:2014oea}.
There are two differences from the round $S^4$ partition function.
One is one-loop determinant \cite{Hama:2012bg,Nosaka:2013cpa}:
\begin{\eqa}
Z_{\rm 1loop} (a)
= \frac{\prod_{\alpha \in \Delta_+} \Upsilon (i a\cdot\alpha ) \Upsilon (-i a\cdot\alpha ) /(\alpha\cdot a)^2 }
{\prod_{m=1}^{N_f} \prod_{\rho_m \in \mathbf{R_m}} \Upsilon \left( i a\cdot\rho +\frac{Q}{2} \right)  } ,
\end{\eqa}
where $Q=b+b^{-1}$ and
\begin{\eq}
\Upsilon (x) 
=\prod_{m_1 ,m_2 \geq 0} (m_1 b +m_2 b^{-1} +x)(m_1 b +m_2 b^{-1} +Q-x) .
\end{\eq}
The other is values of $\Omega$-background parameters 
in $Z_{\rm Nek}$,
which are $\epsilon_1 =b ,\epsilon_2 =b^{-1}$.

Repeating the argument for $Z_{S^4}$,
we can always rewrite $Z_{S_b^4}$ as
the Laplace transformation as in \eqref{eq:Laplace}
and show that
$f^{(k_1 ,\cdots ,k_n )} (t)$ for $Z_{S_b^4}$
is Borel transformation
of original perturbative series.
However, 
pole structure of the Borel transformation
is slightly more involved.
This depends on $b$
and
we have poles in real positive axis
when $m_1 b+m_2 b^{-1}$ ($m_{1,2}\in\mathbb{Z}$) can be purely imaginary 
\footnote{
$b$ is real for ellipsoid \cite{Hama:2012bg}
while $b$ can be complex in the setup of \cite{Nosaka:2013cpa}.
}.
Because
the partition function for this region is ill-defined,
we conclude that
$Z_{S_b^4}$
gives Borel summable perturbative series
when it is well-defined.

%%%%%%%%%%%%%%%%%%%%%%%%%%%%%%%%%%%%%%%%%
%%%%%%%%%%%%%%%%%%%%%%%%%%%%%%%%%%%%%%%%%
%%%%%%%%%%%%%%%%%%%%%%%%%%%%%%%%%%%%%%%%%
\section{5d $\mathcal{N}=1$ theory}
\label{sec:5d}
%%%%%%%%%%%%%%%%%%%%%%%%%%%%%%%%%%%%%%%%%
%%%%%%%%%%%%%%%%%%%%%%%%%%%%%%%%%%%%%%%%%
%%%%%%%%%%%%%%%%%%%%%%%%%%%%%%%%%%%%%%%%%
We also study perturbative series of 5d $\mathcal{N}=1$ SUSY theory.
First
we study squashed $S^5$ partition function
with squashing parameters $(\phi_1 ,\phi_2 ,\phi_3 )$,
which has a simple relation to
SUSY Renyi entropy \cite{Alday:2014fsa}.
We can show 
Borel summability of perturbative series
in a quite parallel way to the 4d case.
The $S^5$ partition function
can be computed by localization \cite{Kim:2012qf}
and this takes the form of \eqref{eq:ZS4}.
While classical part is the same 
up to redefinition of $g$ \footnote{
When we have Chern-Simons (CS) term,
there is ${\rm tr}a^3$ term
but this does not disturb our argument.
Although it would be interesting to study
perturbation by (inverse of) CS level
as in 3d CS matter theory,
this is beyond scope of this paper.
},
one-loop part is given by
\begin{\eq}
 Z_{\rm 1loop}(a) 
= \frac{\prod_{\alpha \in {\rm root} } 
   S_3 (-i\alpha\cdot a ;\vec{\omega} ) /(\alpha\cdot a) }
   {\prod_{m=1}^{N_f} \prod_{\rho_m \in\mathbf{R_m}} 
  S_3 ( -i\rho_m \cdot a +\frac{\omega_1 +\omega_2 +\omega_3}{2} ;\vec{\omega})} ,
\end{\eq} 
where $\vec{\omega}=(\omega_1 ,\omega_2 ,\omega_3 )$
with $\omega_j =1 +\phi_j $ and
\begin{\eq}
 S_3 (z;\vec{\omega} )
= \prod_{n_1 ,n_2 ,n_3 \geq 0} (\vec{n}\cdot \vec{\omega} +z)    
  \prod_{n_1 ,n_2 ,n_3 \geq 1} (\vec{n}\cdot \vec{\omega} -z) ,    
\end{\eq}
with $\vec{n}\cdot \vec{\omega} =n_1 \omega_1 +n_2 \omega_2 +n_3 \omega_3$.
Instanton contribution for this case is
product of three 5d Nekrasov partition functions
with $\Omega$-background parameters $(\epsilon_1 ,\epsilon_2 )=(\phi_2 -\phi_1 ,\phi_3 -\phi_1 )$,
$(\phi_3 -\phi_2 ,\phi_1 -\phi_2 )$ and $(\phi_1 -\phi_3 ,\phi_2 -\phi_3 )$.
Nekrasov partition function in 5d is not rational function 
but ratio of hyperbolic functions.
As in 4d case,
the 5d Nekrasov partition function
does not have poles in real axis 
unless $m_1 \epsilon_1 +m_2 \epsilon_2$ ($m_{1,2} \in\mathbb{Z}$)
can be purely imaginary.
Since
the $S^5$ partition function is ill-defined
when $\vec{n}\cdot\vec{\omega}+(\omega_1 +\omega_2 +\omega_3 )/2$
and $m_1 \epsilon_1 +m_2 \epsilon_2$ can be purely imaginary,
we do not consider these cases.

Now we can prove Borel summability for the $S^5$ partition function as in 4d.
First by the transformation \eqref{eq:trans},
we can rewrite the partition function 
in the form of Laplace transformation as in \eqref{eq:Laplace}.
A similar argument shows that
the integrand is again Borel transformation.
Then as long as the partition function is well-defined,
the Borel transformation does not have singularity
along positive real axis of Borel plane.
Thus we conclude that
the perturbative series of the $S^5$ partition function is Borel summable
for arbitrary number of instanton sector.

We can also show Borel summability for supersymmetric Wilson loop on round $S^5$.
The Wilson loop of the type \eqref{eq:Wilson} is supersymmetric
when its contour is Hopf fibre at one point of $\mathbb{CP}^2$ base.
The result of localization is
\begin{\eq}
\langle W_{\mathbf{R}}({\rm Hopf\ fibre})  \rangle
=\langle {\rm tr}_{\mathbf{R}} e^{2\pi a}\rangle_{\rm M.M.} .
\end{\eq}
Insertion of this operator
does not spoil our argument as in 4d
and thus perturbative series of the SUSY Wilson loop is also Borel summble.

%%%%%%%%%%%%%%%%%%%%%%%%%%%%%%%%%%%%%%%%%
%%%%%%%%%%%%%%%%%%%%%%%%%%%%%%%%%%%%%%%%%
%%%%%%%%%%%%%%%%%%%%%%%%%%%%%%%%%%%%%%%%%
\section{Discussions}
\label{sec:discussions}
%%%%%%%%%%%%%%%%%%%%%%%%%%%%%%%%%%%%%%%%%
%%%%%%%%%%%%%%%%%%%%%%%%%%%%%%%%%%%%%%%%%
%%%%%%%%%%%%%%%%%%%%%%%%%%%%%%%%%%%%%%%%%
We have studied the weak coupling perturbative series 
in 4d $\mathcal{N}=2$ and 5d $\mathcal{N}=1$ SUSY gauge theories 
with Lagrangians.
We have proven 
Borel summability of the perturbative expansions  
in zero instanton sector
for various observables.
We have also proven Borel summability
in arbitrary number of instanton sector
when we know explicit forms of Nekrasov partition functions.
Thus our result is nontrivially consistent 
with the proposal \cite{Argyres:2012vv} 
on the semiclassical realization of IR renormalons. 

Our result also shows that
the Borel resummation in a fixed number of instanton sector
is exactly the same as the truncation of the full result to this sector.
There are two conceptually important implications of this.
First,
we can obtain the exact results
by summing over the Borel resummation 
in each number of instanton sector.
If this is true for all physical observables in 4d $\mathcal{N}=2$ theories,
then one can define 4d $\mathcal{N}=2$ theories in this way
although we have not proven it.
We leave this for our future problem.

Second,
our result means that
the Borel resummation in the zero instanton sector
{\it does not give} the full result including instanton corrections
and 
perturbative series in each number of instanton sector
is ``isolated" in some sense.
Thus perturbative data in different numbers of instanton sector
do not mix with each other.
This feature was already pointed out in \cite{Aniceto:2014hoa}
for $SU(2)$ and $U(2)$ cases.
Our result says that
this is common for quite general 4d $\mathcal{N}=2$ and
5d $\mathcal{N}=1$ theories.
Note also that
this feature itself
was observed long time ago 
in Seiberg-Witten prepotential \cite{Seiberg:1994rs},
which receives only one-loop perturbative corrections but has instanton corrections.
A significant difference from our examples is that
the perturbative correction in the prepotential is trivially Borel summable 
while our examples generally have factorially divergent perturbative expansions,
whose Borel summabilities were a priori nontrivial.   
Similar behaviors appear also in WKB quantization of the quartic oscillator and
$1/N$-expansion of partition function of ABJM theory on $S^3$ \cite{Grassi:2014cla}, for example (see also \cite{Shifman:2014fra}).
The above feature is different from successful examples of resurgence approach,
where perturbative data in zero instanton sector
is related to ones in non-zero instanton sector.
We shall ask
if this feature is common for less SUSY theories or not.

It is interesting to consider
perturbative series of 't Hooft loop in 4d $\mathcal{N}=2$ theory,
whose localization formula has been obtained in \cite{Gomis:2011pf}.
Apparently it seems more involved
because the 't Hooft loop receives 
corrections from monopoles as well as instantons.

It would be also illuminating to study 
perturbative series of 3d CS matter theories on $S^3$
by inverse of CS level as in \cite{Russo:2012kj,Aniceto:2014hoa}.
We cannot naively apply our argument to these theories
because exponential factor is purely imaginary 
for these theories.
Hence we need to change integral contour
to get usual Laplace transformation
but this change picks up residues from poles of integrand.
Probably we should think of it more carefully.

We close by mentioning that
our result would be closely related to
a connection between planar limit and ``very strong coupled large-$N$ limit"
discussed in \cite{Azeyanagi:2012xj}.
It is attractive 
if one can make it more precise from our viewpoint.

%%%%%%%%%%%%%%%%%%%%%%%%%%%%%%%%%%%%%%%%%%%%%%%%%%
\vspace{.5cm}
\acknowledgments
We thank Zohar Komargodski for helpful discussions
and comments on the draft.
We are grateful to Jorge G.~Russo, Ricardo Schiappa, Yuji Tachikawa
and anonymous Referees of PRL
for useful comments on the draft.

%%%%%%%%%%%%%%%%%%%%%%%%%%%%%%%%%%%%%%%%%%%%%%%%%%%%%%%%%%%%%


\begin{thebibliography}{99}
%\cite{Dyson:1952tj}
\bibitem{Dyson:1952tj} 
  F.~J.~Dyson,
  ``Divergence of perturbation theory in quantum electrodynamics,''
  Phys.\ Rev.\  {\bf 85}, 631 (1952).
  %doi:10.1103/PhysRev.85.631
  %%CITATION = doi:10.1103/PhysRev.85.631;%%

%\cite{'tHooft:1977am}
\bibitem{'tHooft:1977am} 
  G.~'t Hooft,
  ``Can We Make Sense Out of Quantum Chromodynamics?,''
  Subnucl.\ Ser.\  {\bf 15}, 943 (1979).
  %%CITATION = SUSEE,15,943;%%

%\cite{Argyres:2012vv}
\bibitem{Argyres:2012vv} 
  P.~Argyres and M.~Unsal,
  ``A semiclassical realization of infrared renormalons,''
  Phys.\ Rev.\ Lett.\  {\bf 109}, 121601 (2012)
  %doi:10.1103/PhysRevLett.109.121601
  [arXiv:1204.1661 [hep-th]],
  %%CITATION = doi:10.1103/PhysRevLett.109.121601;%%
%\cite{Argyres:2012ka}
%\bibitem{Argyres:2012ka} 
%  P.~C.~Argyres and M.~Unsal,
  ``The semi-classical expansion and resurgence in gauge theories: new perturbative, instanton, bion, and renormalon effects,''
  JHEP {\bf 1208}, 063 (2012)
%  doi:10.1007/JHEP08(2012)063
  [arXiv:1206.1890 [hep-th]].
  %%CITATION = doi:10.1007/JHEP08(2012)063;%%

%\cite{Dunne:2012ae}
\bibitem{Dunne:2012ae} 
  G.~V.~Dunne and M.~Unsal,
  ``Resurgence and Trans-series in Quantum Field Theory: The CP(N-1) Model,''
  JHEP {\bf 1211}, 170 (2012)
  %doi:10.1007/JHEP11(2012)170
  [arXiv:1210.2423 [hep-th]],
  %%CITATION = doi:10.1007/JHEP11(2012)170;%%
%\cite{Dunne:2012zk}
%\bibitem{Dunne:2012zk} 
%  G.~V.~Dunne and M.~Unsal,
  ``Continuity and Resurgence: towards a continuum definition of the $\mathbb{CP}$(N-1) model,''
  Phys.\ Rev.\ D {\bf 87}, 025015 (2013)
  %doi:10.1103/PhysRevD.87.025015
  [arXiv:1210.3646 [hep-th]];
  %%CITATION = doi:10.1103/PhysRevD.87.025015;%%
%\cite{Cherman:2013yfa}
%\bibitem{Cherman:2013yfa} 
  A.~Cherman, D.~Dorigoni, G.~V.~Dunne and M.~Unsal,
  ``Resurgence in Quantum Field Theory: Nonperturbative Effects in the Principal Chiral Model,''
  Phys.\ Rev.\ Lett.\  {\bf 112}, 021601 (2014)
 % doi:10.1103/PhysRevLett.112.021601
  [arXiv:1308.0127 [hep-th]];
  %%CITATION = doi:10.1103/PhysRevLett.112.021601;%%
%\cite{Misumi:2014jua}
%\bibitem{Misumi:2014jua} 
  T.~Misumi, M.~Nitta and N.~Sakai,
  ``Neutral bions in the ${\mathbb C}P^{N-1}$ model,''
  JHEP {\bf 1406}, 164 (2014)
 % doi:10.1007/JHEP06(2014)164
  [arXiv:1404.7225 [hep-th]].
  %%CITATION = doi:10.1007/JHEP06(2014)164;%%


%\cite{Unsal:2007jx}
\bibitem{Unsal:2007jx} 
  M.~Unsal,
  ``Magnetic bion condensation: A New mechanism of confinement and mass gap in four dimensions,''
  Phys.\ Rev.\ D {\bf 80}, 065001 (2009)
 % doi:10.1103/PhysRevD.80.065001
  [arXiv:0709.3269 [hep-th]].
  %%CITATION = doi:10.1103/PhysRevD.80.065001;%%  

%\cite{Bogomolny:1980ur}
\bibitem{Bogomolny:1980ur} 
  E.~B.~Bogomolny,
  ``Calculation Of Instanton - Anti-instanton Contributions In Quantum Mechanics,''
  Phys.\ Lett.\ B {\bf 91}, 431 (1980);
  %doi:10.1016/0370-2693(80)91014-X
  %%CITATION = doi:10.1016/0370-2693(80)91014-X;%%
%\cite{ZinnJustin:1981dx}
%\bibitem{ZinnJustin:1981dx} 
  J.~Zinn-Justin,
  ``Multi - Instanton Contributions in Quantum Mechanics,''
  Nucl.\ Phys.\ B {\bf 192}, 125 (1981).
 % doi:10.1016/0550-3213(81)90197-8
  %%CITATION = doi:10.1016/0550-3213(81)90197-8;%%  

%\cite{Poppitz:2011wy}
\bibitem{Poppitz:2011wy} 
  E.~Poppitz and M.~Unsal,
  ``Seiberg-Witten and 'Polyakov-like' magnetic bion confinements are continuously connected,''
  JHEP {\bf 1107}, 082 (2011)
  %doi:10.1007/JHEP07(2011)082
  [arXiv:1105.3969 [hep-th]].
  %%CITATION = doi:10.1007/JHEP07(2011)082;%%
    
%\cite{Russo:2012kj}
\bibitem{Russo:2012kj} 
  J.~G.~Russo,
  ``A Note on perturbation series in supersymmetric gauge theories,''
  JHEP {\bf 1206}, 038 (2012)
 % doi:10.1007/JHEP06(2012)038
  [arXiv:1203.5061 [hep-th]].
  %%CITATION = doi:10.1007/JHEP06(2012)038;%%


%\cite{Aniceto:2014hoa}
\bibitem{Aniceto:2014hoa} 
  I.~Aniceto, J.~G.~Russo and R.~Schiappa,
  ``Resurgent Analysis of Localizable Observables in Supersymmetric Gauge Theories,''
  JHEP {\bf 1503}, 172 (2015)
 % doi:10.1007/JHEP03(2015)172
  [arXiv:1410.5834 [hep-th]].
  %%CITATION = doi:10.1007/JHEP03(2015)172;%%
 
 
%\cite{Gerchkovitz:2016gxx}
\bibitem{Gerchkovitz:2016gxx} 
  E.~Gerchkovitz, J.~Gomis, N.~Ishtiaque, A.~Karasik, Z.~Komargodski and S.~S.~Pufu,
  ``Correlation Functions of Coulomb Branch Operators,''
  arXiv:1602.05971 [hep-th].
  

%\cite{Baggio:2014sna}
\bibitem{Baggio:2014sna} 
  M.~Baggio, V.~Niarchos and K.~Papadodimas,
  ``Exact correlation functions in $SU(2) \mathcal N=2$ superconformal QCD,''
  Phys.\ Rev.\ Lett.\  {\bf 113}, no. 25, 251601 (2014)
  %doi:10.1103/PhysRevLett.113.251601
  [arXiv:1409.4217 [hep-th]],
  %%CITATION = doi:10.1103/PhysRevLett.113.251601;%%
%\cite{Baggio:2014ioa}
%\bibitem{Baggio:2014ioa} 
%  M.~Baggio, V.~Niarchos and K.~Papadodimas,
  ``tt$^{*}$ equations, localization and exact chiral rings in 4d $ \mathcal{N} $ =2 SCFTs,''
  JHEP {\bf 1502}, 122 (2015)
 % doi:10.1007/JHEP02(2015)122
  [arXiv:1409.4212 [hep-th]].
  %%CITATION = doi:10.1007/JHEP02(2015)122;%%

%\cite{'tHooft:1982tz}
\bibitem{'tHooft:1982tz} 
  G.~'t Hooft,
  ``On the Convergence of Planar Diagram Expansions,''
  Commun.\ Math.\ Phys.\  {\bf 86}, 449 (1982).
  %doi:10.1007/BF01214881
  %%CITATION = doi:10.1007/BF01214881;%%


%\cite{Pestun:2007rz}
\bibitem{Pestun:2007rz} 
  V.~Pestun,
  ``Localization of gauge theory on a four-sphere and supersymmetric Wilson loops,''
  Commun.\ Math.\ Phys.\  {\bf 313}, 71 (2012)
 % doi:10.1007/s00220-012-1485-0
  [arXiv:0712.2824 [hep-th]].
  %%CITATION = doi:10.1007/s00220-012-1485-0;%%


%\cite{Nekrasov:2002qd}
\bibitem{Nekrasov:2002qd} 
  N.~A.~Nekrasov,
  ``Seiberg-Witten prepotential from instanton counting,''
  Adv.\ Theor.\ Math.\ Phys.\  {\bf 7}, no. 5, 831 (2003)
  %doi:10.4310/ATMP.2003.v7.n5.a4
  [hep-th/0206161];
  %%CITATION = doi:10.4310/ATMP.2003.v7.n5.a4;%%
%\cite{Nekrasov:2003rj}
%\bibitem{Nekrasov:2003rj} 
  N.~Nekrasov and A.~Okounkov,
  ``Seiberg-Witten theory and random partitions,''
  Prog.\ Math.\  {\bf 244}, 525 (2006)
 % doi:10.1007/0-8176-4467-9_15
  [hep-th/0306238].
  %%CITATION = doi:10.1007/0-8176-4467-9_15;%%



%\cite{Seiberg:1996bd}
\bibitem{Seiberg:1996bd} 
  N.~Seiberg,
  ``Five-dimensional SUSY field theories, nontrivial fixed points and string dynamics,''
  Phys.\ Lett.\ B {\bf 388}, 753 (1996)
  %doi:10.1016/S0370-2693(96)01215-4
  [hep-th/9608111].
  %%CITATION = doi:10.1016/S0370-2693(96)01215-4;%%

 

%\cite{Alday:2009aq}
\bibitem{Alday:2009aq} 
  L.~F.~Alday, D.~Gaiotto and Y.~Tachikawa,
  ``Liouville Correlation Functions from Four-dimensional Gauge Theories,''
  Lett.\ Math.\ Phys.\  {\bf 91}, 167 (2010)
  %doi:10.1007/s11005-010-0369-5
  [arXiv:0906.3219 [hep-th]].
  %%CITATION = doi:10.1007/s11005-010-0369-5;%%
  
 

%\cite{Correa:2012at}
\bibitem{Correa:2012at} 
  D.~Correa, J.~Henn, J.~Maldacena and A.~Sever,
  ``An exact formula for the radiation of a moving quark in N=4 super Yang Mills,''
  JHEP {\bf 1206}, 048 (2012)
  %doi:10.1007/JHEP06(2012)048
  [arXiv:1202.4455 [hep-th]].
  %%CITATION = doi:10.1007/JHEP06(2012)048;%%

%\cite{Fiol:2015spa}
\bibitem{Fiol:2015spa} 
  B.~Fiol, E.~Gerchkovitz and Z.~Komargodski,
  ``The Exact Bremsstrahlung Function in N=2 Superconformal Field Theories,''
  Phys.\ Rev.\ Lett.\  {\bf 116}, 081601 (2016)
 % doi:10.1103/PhysRevLett.116.081601
  [arXiv:1510.01332 [hep-th]].
  %%CITATION = doi:10.1103/PhysRevLett.116.081601;%%


  

%\cite{Nishioka:2013haa}
\bibitem{Nishioka:2013haa} 
  T.~Nishioka and I.~Yaakov,
  ``Supersymmetric Renyi Entropy,''
  JHEP {\bf 1310}, 155 (2013)
 % doi:10.1007/JHEP10(2013)155
  [arXiv:1306.2958 [hep-th]].
  %%CITATION = doi:10.1007/JHEP10(2013)155;%%
 
%\cite{Crossley:2014oea}
\bibitem{Crossley:2014oea} 
  M.~Crossley, E.~Dyer and J.~Sonner,
  ``Super-Renyi entropy \& Wilson loops for $\mathcal{N}=4$ SYM and their gravity duals,''
  JHEP {\bf 1412}, 001 (2014)
 % doi:10.1007/JHEP12(2014)001
  [arXiv:1409.0542 [hep-th]];
  %%CITATION = doi:10.1007/JHEP12(2014)001;%%
%\cite{Huang:2014pda}
%\bibitem{Huang:2014pda} 
  X.~Huang and Y.~Zhou,
  ``$ \mathcal{N}=4 $ Super-Yang-Mills on conic space as hologram of STU topological black hole,''
  JHEP {\bf 1502}, 068 (2015)
%  doi:10.1007/JHEP02(2015)068
  [arXiv:1408.3393 [hep-th]].
  %%CITATION = doi:10.1007/JHEP02(2015)068;%%
 

%\cite{Hama:2012bg}
\bibitem{Hama:2012bg} 
  N.~Hama and K.~Hosomichi,
  ``Seiberg-Witten Theories on Ellipsoids,''
  JHEP {\bf 1209}, 033 (2012)
  [JHEP {\bf 1210}, 051 (2012)]
  %doi:10.1007/JHEP09(2012)033, 10.1007/JHEP10(2012)051
  [arXiv:1206.6359 [hep-th]].
  %%CITATION = doi:10.1007/JHEP09(2012)033, 10.1007/JHEP10(2012)051;%%
 
%\cite{Nosaka:2013cpa}
\bibitem{Nosaka:2013cpa} 
  T.~Nosaka and S.~Terashima,
  ``Supersymmetric Gauge Theories on a Squashed Four-Sphere,''
  JHEP {\bf 1312}, 001 (2013)
 % doi:10.1007/JHEP12(2013)001
  [arXiv:1310.5939 [hep-th]].
  %%CITATION = doi:10.1007/JHEP12(2013)001;%%

  
%\cite{Alday:2014fsa}
\bibitem{Alday:2014fsa} 
  L.~F.~Alday, P.~Richmond and J.~Sparks,
  ``The holographic supersymmetric Renyi entropy in five dimensions,''
  JHEP {\bf 1502}, 102 (2015)
  %doi:10.1007/JHEP02(2015)102
  [arXiv:1410.0899 [hep-th]];
  %%CITATION = doi:10.1007/JHEP02(2015)102;%%
%\cite{Hama:2014iea}
%\bibitem{Hama:2014iea} 
  N.~Hama, T.~Nishioka and T.~Ugajin,
  ``Supersymmetric Renyi entropy in five dimensions,''
  JHEP {\bf 1412}, 048 (2014)
  %doi:10.1007/JHEP12(2014)048
  [arXiv:1410.2206 [hep-th]].
  %%CITATION = doi:10.1007/JHEP12(2014)048;%%

 
%\cite{Kim:2012qf}
\bibitem{Kim:2012qf} 
  H.~C.~Kim, J.~Kim and S.~Kim,
  ``Instantons on the 5-sphere and M5-branes,''
  arXiv:1211.0144 [hep-th];
  %%CITATION = ARXIV:1211.0144;%%
%\cite{Kallen:2012cs}
%\bibitem{Kallen:2012cs} 
  J.~Kallen and M.~Zabzine,
  ``Twisted supersymmetric 5D Yang-Mills theory and contact geometry,''
  JHEP {\bf 1205}, 125 (2012)
  %doi:10.1007/JHEP05(2012)125
  [arXiv:1202.1956 [hep-th]];
  %%CITATION = doi:10.1007/JHEP05(2012)125;%% 
%\cite{Hosomichi:2012ek}
%\bibitem{Hosomichi:2012ek} 
  K.~Hosomichi, R.~K.~Seong and S.~Terashima,
  ``Supersymmetric Gauge Theories on the Five-Sphere,''
  Nucl.\ Phys.\ B {\bf 865}, 376 (2012)
  %doi:10.1016/j.nuclphysb.2012.08.007
  [arXiv:1203.0371 [hep-th]];
  %%CITATION = doi:10.1016/j.nuclphysb.2012.08.007;%%
%\cite{Kallen:2012va}
%\bibitem{Kallen:2012va} 
  J.~Kallen, J.~Qiu and M.~Zabzine,
  ``The perturbative partition function of supersymmetric 5D Yang-Mills theory with matter on the five-sphere,''
  JHEP {\bf 1208}, 157 (2012)
  %doi:10.1007/JHEP08(2012)157
  [arXiv:1206.6008 [hep-th]];
  %%CITATION = doi:10.1007/JHEP08(2012)157;%%
  %\cite{Kim:2012ava}
%  \bibitem{Kim:2012ava} 
    H.~C.~Kim and S.~Kim,
    %``M5-branes from gauge theories on the 5-sphere,''
    JHEP {\bf 1305}, 144 (2013);
 %   doi:10.1007/JHEP05(2013)144
    [arXiv:1206.6339 [hep-th]].
    %%CITATION = doi:10.1007/JHEP05(2013)144;%%
%\cite{Lockhart:2012vp}
%\bibitem{Lockhart:2012vp} 
  G.~Lockhart and C.~Vafa,
  ``Superconformal Partition Functions and Non-perturbative Topological Strings,''
  arXiv:1210.5909 [hep-th];
  %%CITATION = ARXIV:1210.5909;%%
%\cite{Imamura:2012bm}
%\bibitem{Imamura:2012bm} 
  Y.~Imamura,
  ``Perturbative partition function for squashed $S^5$,''
 PTEP {\bf 2013}, no. 7, 073B01 (2013)
%  doi:10.1093/ptep/ptt044
  [arXiv:1210.6308 [hep-th]].
  %%CITATION = doi:10.1093/ptep/ptt044;%%

%\cite{Seiberg:1994rs}
\bibitem{Seiberg:1994rs} 
  N.~Seiberg and E.~Witten,
  ``Electric - magnetic duality, monopole condensation, and confinement in N=2 supersymmetric Yang-Mills theory,''
  Nucl.\ Phys.\ B {\bf 426}, 19 (1994)
  Erratum: [Nucl.\ Phys.\ B {\bf 430}, 485 (1994)]
 % doi:10.1016/0550-3213(94)90124-4
  [hep-th/9407087],
  %%CITATION = doi:10.1016/0550-3213(94)90124-4;%%
%\cite{Seiberg:1994aj}
%\bibitem{Seiberg:1994aj} 
  %N.~Seiberg and E.~Witten,
  ``Monopoles, duality and chiral symmetry breaking in N=2 supersymmetric QCD,''
  Nucl.\ Phys.\ B {\bf 431}, 484 (1994)
  %doi:10.1016/0550-3213(94)90214-3
  [hep-th/9408099].
  %%CITATION = doi:10.1016/0550-3213(94)90214-3;%%

%\cite{Grassi:2014cla}
\bibitem{Grassi:2014cla} 
  A.~Grassi, M.~Marino and S.~Zakany,
  ``Resumming the string perturbation series,''
  JHEP {\bf 1505}, 038 (2015)
  %doi:10.1007/JHEP05(2015)038
  [arXiv:1405.4214 [hep-th]].
  %%CITATION = doi:10.1007/JHEP05(2015)038;%%

%\cite{Shifman:2014fra}
\bibitem{Shifman:2014fra} 
  M.~Shifman,
  ``Resurgence, operator product expansion, and remarks on renormalons in supersymmetric Yang-Mills theory,''
  J.\ Exp.\ Theor.\ Phys.\  {\bf 120}, no. 3, 386 (2015)
  %doi:10.1134/S1063776115030115
  [arXiv:1411.4004 [hep-th]].
  %%CITATION = doi:10.1134/S1063776115030115;%%
  
%\cite{Gomis:2011pf}
\bibitem{Gomis:2011pf} 
  J.~Gomis, T.~Okuda and V.~Pestun,
  ``Exact Results for 't Hooft Loops in Gauge Theories on $S^4$,''
  JHEP {\bf 1205}, 141 (2012)
  %doi:10.1007/JHEP05(2012)141
  [arXiv:1105.2568 [hep-th]].
  %%CITATION = doi:10.1007/JHEP05(2012)141;%%

%\cite{Azeyanagi:2012xj}
\bibitem{Azeyanagi:2012xj} 
  T.~Azeyanagi, M.~Fujita and M.~Hanada,
  ``From the planar limit to M-theory,''
  Phys.\ Rev.\ Lett.\  {\bf 110}, no. 12, 121601 (2013)
  %doi:10.1103/PhysRevLett.110.121601
  [arXiv:1210.3601 [hep-th]];
  %%CITATION = doi:10.1103/PhysRevLett.110.121601;%%
%\cite{Azeyanagi:2013fla}
%\bibitem{Azeyanagi:2013fla} 
  T.~Azeyanagi, M.~Hanada, M.~Honda, Y.~Matsuo and S.~Shiba,
  ``A new look at instantons and large-N limit,''
  JHEP {\bf 1405}, 008 (2014)
 % doi:10.1007/JHEP05(2014)008
  [arXiv:1307.0809 [hep-th]].
  %%CITATION = doi:10.1007/JHEP05(2014)008;%%
  
\end{thebibliography}
\end{document}